\documentclass{statsoc}

\usepackage{epsfig}                 
\usepackage{rotating}               

\newcommand{\SiII}{\ensuremath{\mathrm{Si^{\raisebox{0.1pt}{+}}}}}
\newcommand{\norm}[1]{\raisebox{1pt}{\ensuremath{|\!|}}#1%
                      \raisebox{1pt}{\ensuremath{|\!|}}}

\newcommand{\mbold}[1]{\mbox{\boldmath$ #1 $\unboldmath}}

\newcommand{\bbeta}{\mbold{\beta}}

\newcommand{\bmu}{\mbold{\mu}}

\newcommand{\by}{\mathbf{y}}

\newcommand{\bD}{\mathbf{D}}

\newcommand{\bI}{\mathbf{I}}

\newcommand{\bU}{\mathbf{U}}
\newcommand{\bV}{\mathbf{V}}
\newcommand{\bW}{\mathbf{W}}

\newcommand{\bY}{\mathbf{Y}}

\newcommand{\edit}[1]{#1}

\title[Estimating Uncertainties of Stellar Oscillation-Related
Parameters]{Estimating  Stellar Oscillation-Related
Parameters and Their Uncertainties with the Moment Method}

\author[J. De Ridder, G. Molenberghs and C. Aerts]{Joris De Ridder}
\address{Institute of Astronomy - Katholieke Universiteit Leuven,
         B-3001 Leuven, Belgium}
\email{joris@ster.kuleuven.ac.be}

\author[J. De Ridder, G. Molenberghs and C. Aerts]{Geert Molenberghs}
\address{Center for Statistics - Limburgs Universitair Centrum,
         B-3590 Diepenbeek, Belgium}
\email{geert.molenberghs@luc.ac.be}

\author[J. De Ridder, G. Molenberghs and C. Aerts]{Conny Aerts}
\address{Institute of Astronomy - Katholieke Universiteit Leuven,
         B-3001 Leuven, Belgium}
\email{conny@ster.kuleuven.ac.be}

\begin{document}


\begin{abstract}
The moment method is a well known mode identification technique in 
asteroseismology (where `mode' is to be understood in an astronomical rather than in a statistical sense),  which uses a time series of the first 3 moments of 
a spectral line to estimate the discrete {oscillation} mode 
parameters $\ell$ and $m$.
The method, contrary to many other mode identification techniques, also
provides estimates of other important continuous parameters such as the inclination
angle $\alpha$, and the rotational velocity $v_e$. We developed a statistical
formalism for the moment method based on so-called generalized estimating
equations (GEE). This formalism allows the estimation of the uncertainty of the
continuous parameters taking into account that the different moments
of a line profile are correlated and that the
uncertainty of the observed moments also depends on the model parameters.
Furthermore, we set up a procedure to take into account the mode 
uncertainty, i.e., the fact that often several modes $(\ell,m)$ can adequately 
describe the data. 
{We also introduce a new lack of fit function
which works at least as well as a previous discriminant function, and which 
in addition allows us to identify the sign of the azimuthal order $m$.} 
We applied our method to the star HD181558 {using several numerical methods}, 
from which we learned that numerically solving the estimating equations
is an intensive task. {We report on the numerical results, from which
we gain insight in the statistical uncertainties of the physical parameters
involved in the moment method.}
\keywords{Generalized estimating equations, time series, sandwich estimator,
astrostatistics, discriminant function}
\end{abstract}


\section{Introduction}
\label{introduction}
Stars consist of a number of gas layers with different temperatures, pressures, and chemical compositions. During their sojourn on the main-sequence, i.e., when
they transform hydrogen into helium, some stars are subject to oscillations which in turn provide astronomers with a wealth of information about the stellar interior. This is the subject of asteroseismology. Such oscillations typically exhibit multiple frequencies and
 manifest
themselves at the surface of the star through variations in brightness, 
temperature, and surface velocity; some of these are observable.
A star can oscillate in one or more of its ``natural'' frequencies determined by the internal structure of the star. With suitable
inversion techniques it is possible to use the observed frequencies to 
derive information about this internal structure.

To do so, however, the characteristics of the oscillations need to be considered
first. That is, a \emph{mode identification}\/ (note that `mode' is an astronomical term) has to be carried out, in which
one estimates the parameters characterising the oscillations from 
observational data. There are few mode identification techniques, and the
properties of their estimators are rarely studied. Statistical 
uncertainties of the estimates, for example, are never reported.
Nevertheless, from an astrophysical point of view, such uncertainties are important
because wrong mode identifications can bias inversion techniques.
It is therefore necessary to know a priori the extent of possible errors in the estimates. 

In this paper, we study the statistical properties of
one particular mode identification technique,  the so-called 
\emph{moment method}. For examples of applications of this method we refer
to, e.g., Aerts et al.~(1998), Uytterhoeven et al.~(2001), Aerts \& Kaye (2001),
and Chadid et al.\ (2001).


\section{Astrophysical Background}

As in any inferential method, the moment method uses a theoretical model
to describe the observations. To understand the statistically relevant 
properties of this theoretical model and its parameters, we first  
 briefly discuss some of the physics of stellar oscillations and how 
they are observed.

\begin{figure}[h]
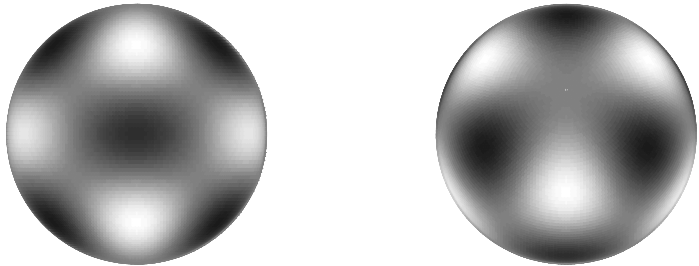

\centering
\begin{turn}{270}
\epsfig{file=bol1.eps,height=0.25\textwidth}
\end{turn}
\hspace*{20mm}
\begin{turn}{270}
\epsfig{file=bol2.eps,height=0.25\textwidth}
\end{turn}
\caption{A diagrammatic illustration of {an orthographic planar
projection of} the surface of an oscillating stars
for the mode $(\ell,m) = (5,3)$. In the left picture we look at the
equator, and in the right picture we look almost at the pole of the star.   
\label{Ylm}}
\end{figure}

Figure~\ref{Ylm} gives a diagrammatic illustration of an orthographic
planar projection of the surface of an oscillating star, i.e., parallel with the line of sight.
For our application, the most important aspect of stellar 
oscillation is the surface velocity. The lighter parts of the stellar surface
have an inward velocity while the darker parts have an outward velocity.
The figure is only a snapshot: the star varies periodically and half an 
oscillation cycle later the situation is reversed with the lighter parts
moving outward and the darker parts moving inward. For slowly-rotating 
oscillating stars, each of the oscillation modes can be described with a 
spherical harmonic $Y_{\ell}^m$ (which is  actually the basis of our illustration in Figure 
\ref{Ylm}), where $\ell$ is the total number of nodal lines and $m$ is the number of nodal lines perpendicular to the equator.  In reality, the motion of a surface element is more complex because it moves horizontally as well as vertically.

In terms of model parameters we have to estimate 3 unknown parameters 
per oscillation mode: 2 discrete parameters and 1 continuous parameter. 
The 2 discrete parameters are the mode numbers $\ell$ and $m$ 
of the spherical harmonic, which describe the configuration 
of the inward and outward going regions. To describe the 3-dimensional 
motion of the stellar matter, only one parameter is needed: the 
amplitude $v_p$ of the vertical motion, since there is
a theoretical linear relation between the amplitude of the vertical
motion and the amplitude of the horizontal motion. To compute the constant
of proportion $K$, however, the mass and the radius of the star are required
and these quantities are often not very accurately known. 
Nevertheless, in what follows we will assume, as a first approximation, 
that $K$ is known, to considerably simplify the treatment. 

A further continuous parameter related to the oscillation is the oscillation period $P$. However, 
for good datasets, this oscillation period can often be quite accurately 
determined from the data with other methods so that it is usually regarded as known.

In the model, 2 additional unknown parameters not connected to the 
oscillations are present. A first one is the rotational velocity at the 
equator of the star, usually denoted as $v_e$. The second one is the 
inclination angle $\alpha$ under which we observe the star. This is illustrated 
in Figure \ref{Ylm}. Both pictures show the same $Y_{\ell}^m$, but on the 
left hand side we are looking on the equator, while on the right hand side
we are looking almost on the pole. Clearly, $\alpha$ has a large impact on how 
the surface velocity field is observed.  
 
\begin{figure}[h]
\centering
\begin{turn}{270}
\epsfig{file=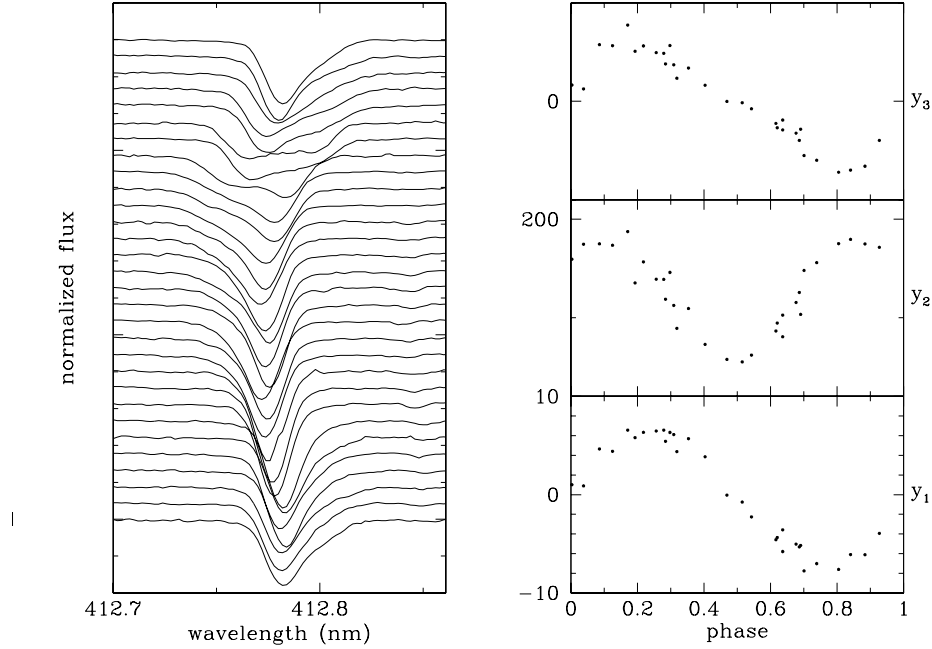,height=0.9\textwidth}
\end{turn}
\caption{In the left panel, a time series of Si$^{+}$ (412.805 nm) absorption
lines of the non-radially oscillating star HD181558 is shown. 
The line profiles are `sorted' to cover an entire
oscillation cycle of the dominant mode, which has a period of about 29h42m.
Each of the line profiles is vertically shifted to obtain a clear visual
effect. In the right panels, the first moment $y_1$ (in km/s), the second
moment $y_2$ (in km$^2$/s$^2$) and the third moment 
$y_3$ (in km$^3$/s$^3$) of all line profiles are shown as
a function of the oscillation phase $\phi$.
\label{hd181558_spec}}
\end{figure}

A last unknown model parameter is specifically related to the kind of 
observational data we use. In the case of the moment method it concerns
high-resolution spectroscopic data. The gathered star light is
decomposed into its colours so that a detailed spectrum can be constructed, 
i.e., received light flux as a function of the wavelength of the light. 
At certain wavelengths, such a spectrum contains absorption lines where the 
light has been partially blocked by certain chemical elements at the surface
of the star. An example of the \SiII\ absorption line at $\lambda$ = 412.805 nm
for the non-radially oscillating star HD181558 is shown in the left hand panel 
of Figure \ref{hd181558_spec}. Here, an observational time series of $N=30$ 
high-quality spectra gathered by De Cat and Aerts~(2002) is shown.
The oscillations in the star cause the absorption line to change its position
and shape in time. Precisely these line profile variations are used to
estimate the parameters mentioned above. To model them, another unknown
parameter is needed, denoted by $\sigma$, which is related to the width
the line profile would have in the absence of pulsation. From an 
astrophysical point of view this is an unimportant nuisance parameter. 
For convenience,
Table~\ref{paramsummary} summarizes all unknown model parameters mentioned above, their 
meaning and their physical range.
\begin{table}
\caption{A summary of all unknown relevant model parameters, with their meaning and their 
physical range. The notation $\bbeta = (\beta_1, \beta_2, \beta_3, \beta_4)'$ for 
the continuous parameters will be introduced and used in Section \ref{currentstatus}.
\label{paramsummary}}
\centering
\fbox{%
\renewcommand{\arraystretch}{1.5}
\begin{tabular}{l@{\qquad}l@{\qquad}l}
Parameter & Meaning & Physical Range \\
\hline
$\ell$             & \parbox[t]{17em}{Degree of the spherical harmonic $Y_{\ell}^m$} & $\{0,1,2,\cdots\}$ \\
$m$                & \parbox[t]{17em}{Azimuthal order of the spherical harmonic $Y_{\ell}^m$} & $\{-\ell,\cdots,0,\cdots,+\ell\}$ \\
$v_p = \beta_1$    & \parbox[t]{17em}{Velocity amplitude of the oscillation} & $\ge 0$ \\
$\sigma = \beta_2$ & \parbox[t]{17em}{Width of line profile in absence of pulsation and rotation (nuisance)} & $\ge 0$  \\
$v_e = \beta_3$    & \parbox[t]{17em}{Equatorial rotational velocity} & $> 0$  \\
$\alpha = \beta_4$      & \parbox[t]{17em}{Inclination angle of the star} & $[0^{\circ},360^{\circ})$ \\
\end{tabular}}
\end{table}

Modeling the line profiles themselves turns out to be \emph{very}
computationally expensive. That is why Balona (1986) devised the moment 
method, which replaces each line profile by  the first, 
second, and third moment denoted by $y_1$, $y_2$, 
and $y_3$ respectively. These quantities are \edit{\emph{measures}} for
the average position, the square of the width and the skewness 
of the line profile. Precisely,
\begin{equation}
y_n(\phi)=\frac{
\int_{-\infty}^{+\infty}[1-p(\phi,\lambda)]\lambda^n\,d\lambda
}
{
\int_{-\infty}^{+\infty}[1-p(\phi,\lambda)]\,d\lambda
},
\label{y123}
\end{equation}
where $p(\phi,\lambda)$ is the line profile function at phase $\phi$ and for wavelength $\lambda$. For each $\phi$, there is a separate line profile in the left hand panel of Figure~\ref{hd181558_spec}, leading to a point for each of the right hand side sub-panels, corresponding to $n=1,2,3$.
In practice, no higher moments  are considered since these are often noisy and unduly
complicate the calculations. One commonly expresses the 
moments in (km/s)$^n$, by transforming the wavelength $\lambda$ in (\ref{y123}) to a velocity using the Doppler transformation formula:
$$v=c\frac{\lambda-\lambda_0}{\lambda_0},$$
where $c$ is the speed of light.
The moments $y_n(\phi)$ can be expressed in terms of time $t$ as well, where $\phi$ is defined as $t\mbox{mod}P/P$, with `mod' standing for the decimal part and $P$ the oscillation period. 
A time series of theoretical moments can be computed much faster than
one of theoretical line profiles. The nuisance parameter $\sigma$, 
however, remains. In the right hand panels of Figure~\ref{hd181558_spec} 
we show a time series of the three moments for the star HD181558. The computation of such moments from spectral line profiles take the form of intensity-weighted  sums, sums of squares, and sums of cubes. 

Fifteen years after the introduction of the moment method,
this mode identification technique is still very relevant
(see the recent references given in Section~\ref{introduction}). Indeed, the effort required for direct computation of time series of 
line profiles is currently still too computationally demanding to be useful for mode 
identification, even under simplifying assumptions concerning the shape of the absorption line.

A theoretical moment at one point of time is computed by integrating over the 
contributions of all points on the visible stellar surface.
Closed-form expressions for the moments exist (Aerts et al., 1992), but they are quite lengthy and of little 
practical use to computing derivatives, such as in Section~\ref{section5}.
We opt for different, computationally more advantageous, expressions, 
which involve an integration with bounds
depending on the inclination angle $\alpha$.


\section{Current Statistical Status}
\label{currentstatus}
The moment method is a multi-response problem where a time
series of 3 responses is used to extract 6 (i.e., 2 discrete and 4 
continuous) parameters. In what follows we will use the notation
$$\bY_i \equiv  (y_1(t_i),\ y_2(t_i),\ y_3(t_i))'
$$
and
$$
\bmu_i \equiv (\mu_1(t_i, \ell,m,\bbeta),\ \mu_2(t_i, \ell,m,\bbeta),\ 
                \mu_3(t_i, \ell,m,\bbeta))'
$$
for the first three observed and theoretical moments respectively, at time 
point $t_i \ (i = 1,\cdots,n)$, where  $\bbeta \equiv (v_p,\sigma,v_e,\alpha)'$.

It is important to understand how the moment method is currently used.
Theoretically, it can be shown (Aerts et al., 1992)
that for a monoperiodic star, the time dependence of the moments takes the 
following form:
\begin{equation}
\left.\begin{array}{rcl}
\mu_1 & = & a_1\ \sin(2\pi\nu t + \kappa_1),\\
\mu_2 & = & b_0 + b_1\ \sin(2\pi\nu t + \delta_1)
                               + b_2\ \sin(4\pi\nu t + \delta_2), \\
\mu_3 & = &   c_1 \ \sin(2\pi\nu t + \gamma_1)
                          +  c_2 \ \sin(4\pi\nu t + \gamma_2)
                          +  c_3 \ \sin(6\pi\nu t + \gamma_3),
\end{array}
\right\}
\label{moments}
\end{equation}
where $\nu$ is the oscillation frequency. The phases $\alpha_1$, $\delta_i$,
$\gamma_j$ are constants, while the \edit{positive} amplitudes 
$a_1$, $b_i$, $c_j$ depend on the parameters $(\ell,m,\bbeta$). 
A \emph{discriminant} $\Gamma_l^m(\bbeta)$ is
constructed to estimate these parameters by comparing the observed amplitudes
with their theoretical counterparts:
\begin{equation}
\label{discriminant}
\Gamma_{\ell}^m =
\left\{\bigg(f_{\tilde{a}_1}\ |\tilde{a}_1 - a_1|\bigg)^2
+ \sum\limits_{i=0}^2 \left(f_{\tilde{b}_i}\ \sqrt{|\tilde{b}_i - b_i|}\right)^2
+ \sum\limits_{i=1}^3 \left(f_{\tilde{c}_i}\ \sqrt[3]{\rule{0pt}{9pt}%
|\tilde{c}_i - c_i|}\right)^2\right\}^{1/2} ,
\end{equation}
where the tilde denotes observed quantities, and where the weights $f$ 
are introduced to incorporate the estimated standard errors 
$\Delta\tilde{a}_1$, $\Delta\tilde{b}_i$ and $\Delta\tilde{c}_j$ of the
corresponding observed amplitudes:
\begin{displaymath}
f_{\tilde{a}_1} \equiv W^{-1}\ \frac{\tilde{a}_1}{\Delta\tilde{a}_1},
\ \ \ \ \ \ \ \ \ \ \
f_{\tilde{b}_i} \equiv W^{-1}\ \frac{\tilde{b}_i}{\Delta\tilde{b}_i},
\ \ \ \ \ \ \ \ \ \ \
f_{\tilde{c}_i} \equiv W^{-1}\ \frac{\tilde{c}_i}{\Delta\tilde{c}_i},
\end{displaymath}
\begin{displaymath}
W \equiv \frac{\tilde{a}_1}{\Delta\tilde{a}_1}
       + \frac{\tilde{b}_0}{\Delta\tilde{b}_0}
       + \frac{\tilde{b}_1}{\Delta\tilde{b}_1}
       + \frac{\tilde{b}_2}{\Delta\tilde{b}_2}
       + \frac{\tilde{c}_1}{\Delta\tilde{c}_1}
       + \frac{\tilde{c}_2}{\Delta\tilde{c}_2}
       + \frac{\tilde{c}_3}{\Delta\tilde{c}_3}
\end{displaymath}
(Aerts 1996). The form of  $\Gamma_{\ell}^m$  in
(\ref{discriminant}) prevents the third moment $y_3$ with its large values from dominating
the first moment
$y_1$, but has the disadvantage that it cannot discriminate the sign of the
mode number $m$. The parameters are estimated by searching for the minimum of 
$\Gamma_{\ell}^m$ in a rectangular grid in the parameter space. For the
continuous parameters, it is hoped for that the grid is fine enough in order
not to miss the global minimum. Finally, a table is produced with 
the top 5 or 6 best fitting $(\ell,m,\bbeta)$ parameter sets. 

The best strategy to obtain the final estimate for $(\ell,m)$ and 
$\bbeta$ together with their uncertainties, is currently open to debate. 
Despite the usefulness of the table with the best parameter sets, there is no estimate of the uncertainties of the parameters obtained 
with the moment method, because of severe theoretical and computational complexity. This paper takes an important first step towards estimating the uncertainties of the continuous
parameters $\bbeta$. Estimating the uncertainties of the discrete parameters 
$\ell$ and $m$ is an \edit{even} more challenging problem, and will be left 
for future research. 

Even for a given $(\ell,m)$ value, it is currently unknown how precise the
continuous parameters are estimated. For example, is the uncertainty in the 
inclination angle as small as 5$^{\circ}$, or is perhaps 30$^{\circ}$ a more
typical value? Moreover, very often several $(\ell,m)$ pairs
give almost equally good fits. The question is raised as to how should 
we take this into account for our best estimate of $\bbeta$ and its 
uncertainty? In what follows, we will try to answer these questions.


\section{New Statistical Approach}

\label{GEE_theory}
We consider a new estimating method which  
produces both point and interval estimates.

\edit{We first note that} the three responses $y_1$, $y_2$ and
$y_3$ are dependent, and that their covariance matrix $\bV$ is unknown.
\edit{Formulating a statistical model for the noise on the moments
is non-trivial as it would involve a model for both the instrumental and
the atmospheric noise. In addition, we note that the relation between the 
coefficients $a_1$, $b_i$, $c_i$ in (\ref{moments}), and the 
parameters $\bbeta$ is non-linear thereby preventing the easy computation of a 
Jacobian matrix. Estimating $\bbeta$ and its covariance matrix 
${\rm Cov}[\bbeta]$ using a simple variable transformation technique is 
therefore not possible.}

A first \edit{alternative} is the least squares method.
\edit{Although multi-response least-squares estimation has been used before
to deal with correlated responses where the covariance matrix has to be
estimated, Seber and Wild (1989) show that this technique should \emph{not}
be used if the covariance matrix $\bV$ depends on the parameters $\bbeta$, 
as is the case here.}
For example, the uncertainty in the first moment of a line profile ($y_1$) can
be estimated with the second moment ($y_2$), and the latter depends on 
$\ell$, $m$, and $\bbeta$. Or, with an astrophysical example, the faster the
star rotates (larger $v_e$) the broader and flatter the line profile, and
the less precision with which we know the position or the first moment of the line
profile. \edit{To avoid confusion, we stress that our argument in the example
above is not that the uncertainty of the first moment $y_1$ depends on the 
uncertainty of the second moment $y_2$, but that the uncertainty of the first 
moment $y_1$ always depends on the second moment itself of which we know that 
it depends in turn on the parameters $\bbeta$.}

\edit{We must therefore conclude that} in the case of the moment method, 
minimizing the weighted sum of squares is not appropriate, \edit{regardless
of how $\bV$ is estimated, because it will not yield a consistent estimate
of the parameters $\bbeta$ and it will reduce the efficiency of the
estimator (Seber and Wild, 1989).}

The \emph{generalized estimating equations} (GEE) methodology, as developed 
by Liang and Zeger (1986), is better suited for the purpose of the 
moment method.
We recall that this method does not assume a particular \emph{joint} 
probability density for the responses $y_1$, $y_2$ and $y_3$, 
nor that they are i.i.d. The theory does not assume that the 
theoretical model is linear in its parameters, while the covariance 
matrix $\bV$ of the responses does not need to be completely specified. 
The method \emph{does} assume, however, that the different observations 
$\bY_i \ (i = 1,\cdots,n)$ are independent, that a working approximation of 
the covariance matrix of the responses is available, and that the 
expectation values $E[\bY_i] \equiv \bmu_i(\ell,m,\bbeta) \ (i =1,\cdots,n)$ 
are correctly specified.

We therefore use GEE to estimate the uncertainties of  the continuous parameters
$\bbeta$. We recall that in the GEE method, the parameters are estimated by 
locating the root of the quasi-score function $\bU(\bbeta)$:
\begin{equation}
\label{U_function}
\bU(\bbeta) \equiv
  \sum\limits_{i=1}^N \bD^t_i \cdot\bW_i^{-1}\cdot(\bY_i - \bmu_i),
\end{equation}
where $N$ is the size of the time series. The $3 \times 4$ matrix 
$\bD= \partial \bmu/\partial \bbeta^t$,
and the $3 \times 3$ symmetric matrix $\bW_i$ is a working approximation 
of the true covariance matrix $\bV_i$ of the quantities $\bY_i$:
\begin{equation}
\bV_i \equiv E[(\bY_i - \bmu_i(\bbeta)) (\bY_i - \bmu_i(\bbeta))^t],
\end{equation}
where $\bbeta$ are the true (but unknown) parameters. It can be shown 
(e.g., Liang and Zeger, 1986; Zeger and Liang 1986; Diggle et al.~2002) that 
the root $\hat{\bbeta}$ is a consistent and asymptotically normal estimate 
of the true $\bbeta$, with sandwich covariance matrix
\begin{equation}
\label{cov_beta}
{\rm Cov}[\hat{\bbeta}] = \mathbf{I}_0^{-1} \ \mathbf{I}_1 \ \mathbf{I}_0^{-1},
\end{equation}
where 
$$\mathbf{I}_0 \equiv 
    -E\left[\frac{\partial\bU(\bbeta)}{\partial\bbeta}\right]
  = \sum\limits_{i=1}^N   \bD_i^t \ \bW_i^{-1} \ \bD_i 
$$
and
$$
\mathbf{I}_1  \equiv  {\rm Cov}[\bU(\bbeta)]
  =  \sum\limits_{i=1}^N \bD_i^t \ \bW_i^{-1} \ \bV_i \ \bW_i^{-1} \ \bD_i.
$$
The unknown covariance matrices $\bV_i$ in the expression for 
$\mathbf{I}_1$  are estimated by
$$
\hat{\bV_i} =
  (\bY_i - \bmu_i(\hat{\bbeta})) \cdot (\bY_i - \bmu_i(\hat{\bbeta}))^t .
$$
The so-called sandwich estimator in (\ref{cov_beta}) is robust against 
misspecification of the covariance matrix of the responses. 

For the working approximation $\bW_i$ for the covariance matrix $\bV_i$, we
suggest the following idea, where we estimate the uncertainty of the 
first
three moments of the line profile with the higher moments, as is sometimes
done with the moments of a probability distribution function. 
Consider the mirror image 
$\zeta(\phi,v) = 1 - p(\phi,v)$ of the spectral line $p(\phi,v)$  as a distribution 
function for the velocity $v$, and compute, for a given time $t_i$:
\begin{eqnarray*}
W_{i,rs}\equiv W_{rs} & = & E[(y_r - \mu_r) (y_s - \mu_s)] \nonumber \\
       & = & E\left[\left(\frac{\sum \zeta_j \ v_j^r}{\sum \zeta_j}
                          - \mu_r\right)
                    \left(\frac{\sum \zeta_j \ v_j^s}{\sum \zeta_j}
                          - \mu_s\right)\right] \nonumber \\
       & = & \frac{1}{\left(\sum \zeta_j \right)^2} \
             E\left[\sum\limits_j \zeta_j (v_j^r - \mu_r) \cdot
                     \sum\limits_j \zeta_j (v_j^s - \mu_s) \right] \nonumber \\
       & = & \frac{1}{\left(\sum \zeta_j \right)^2} \
             \sum\limits_j \zeta_j^2 \ (E[v_j^r \ v_j^s] - \mu_r \ \mu_s)
               \nonumber \\
       & = & \Gamma \cdot (\mu_{r+s} - \mu_r \ \mu_s),
\end{eqnarray*}
where the sum over the index $j$ runs over all the velocity points (pixels) of the spectral line,
and where we define
$$
\Gamma \equiv \sum_j \zeta_j^2 \left/ \bigg(\sum_j \zeta_j\bigg)^2\right..
$$
Here, we assume that the different observed points of the line profile are 
uncorrelated. \edit{The extra factor $\Gamma$ 
appears because, contrary to probability distribution functions, line profiles are not normalized in area.}
\edit{Note that we use the higher \emph{theoretical} moments and not 
the observational ones because, as mentioned before, 
the latter are often too noisy.
It is difficult to assess the influence of the working approximation
$\bW$ on the final uncertainties on $\bbeta$, but we refer to 
Diggle et al.~(2002) where it is shown that the sandwich estimator 
(\ref{cov_beta}) for the covariance matrix of $\bbeta$ is quite robust 
against misspecification of $\bV$.}

Having derived an estimator for $\bbeta$ and its uncertainty, given
an $(\ell,m)$ pair, we should take into account that we do not actually
know the correct $(\ell,m)$ values. \edit{If $\ell$ and $m$ were continuous
parameters, we would have a total of 6 continuous parameters for which we
would have liked to compute a 6-dimensional confidence region. As $\ell$ and
$m$ are discrete, however, it is notoriously hard to find an equivalent 
``confidence region''. We remind that it is current practice simply to take
 the $\bbeta$ values of the best-fitting $(\ell,m)$ pair with no error estimate
at all. As a first alternative, we propose to ``weight''} each
mode $(\ell,m)$ with a lack-of-fit function. The best guess for
both $\bbeta$ and its uncertainty is then computed with a weighted
mean over all relevant modes $(\ell,m)$. The entire estimation procedure 
can be summarized as follows:
\begin{enumerate}
\item Specify a set of pairs of the degree $\ell$ and the azimuthal number
      $m$: $\{(\ell_j, m_j)\}$.
\item For each of the pairs $(\ell_j, m_j)$, solve the quasi-score
      equations and estimate the continuous parameters $\hat{\bbeta}_j$ and
      their covariance matrix ${\rm Cov}[\hat{\bbeta}_j]$.
\item Compute for each of the modes $(\ell_j, m_j)$, the lack-of-fit parameter
      $G^2_j$ which indicates how well the theoretical moments
      $\bmu(\hat{\bbeta})$ fit the observed moments $\by$:
      \begin{equation}
      \label{G_weight}
      G^2_j = \sum\limits_{k=1}^3\sum\limits_{i=1}^N
        \frac{(y_k(t_i) - \mu_k(\hat{\bbeta_j}, t_i))^2}{
              \mu_{2k}(\hat{\bbeta_j}, t_i) - \mu_k^2(\hat{\bbeta_j}, t_i)}.
      \end{equation}
\item The best estimate for the degree and the azimuthal number
      $(\tilde{\ell}, \tilde{m})$ is the $(\ell_j, m_j)$ that has the lowest
      lack-of-fit $G^2_j$. The corresponding best estimate for the continuous
      parameters $\tilde{\bbeta}$ can be computed with
      \begin{equation}
      \label{tildebeta}
      \tilde{\bbeta} = \frac{\sum\limits_{\{(\ell_j,m_j)\}}
                              \hat{\bbeta}_j \ G^{-2}_j}{
                            \sum\limits_{\{(\ell_j,m_j)\}} G^{-2}_j}
      \end{equation}
      and the corresponding covariance matrix is the sum of the
      intra-mode variance and the inter-mode variance:
      \begin{equation}
      \label{cov_tildebeta}
      {\rm Cov}[\tilde{\bbeta}] = \frac{\sum\limits_{\{(\ell_j,m_j)\}}
                              {\rm Cov}[\hat{\bbeta_j}] \ G^{-2}_j}{
                             \sum\limits_{\{(\ell_j,m_j)\}} G^{-2}_j}
                       +    \frac{\sum\limits_{\{(\ell_j,m_j)\}}
                              (\tilde{\bbeta} - \hat{\bbeta}_j) \cdot
                              (\tilde{\bbeta} - \hat{\bbeta}_j)^t \ G^{-2}_j}{
                             \sum\limits_{\{(\ell_j,m_j)\}} G^{-2}_j}.
      \end{equation}
\end{enumerate}
For both practical and astrophysical reasons, only modes with a degree $\ell$
up to a certain limit (e.g. $\ell \le 4$) are considered.

In the following section this estimation procedure is applied to a dataset 
of the star HD181558.


\section{Application to HD181558}
\label{section5}
HD181558 belongs to the class of the Slowly Pulsating B stars (SPBs).
Although the star is multi-periodic (De Cat and Aerts 2002) 
it has a very dominant (in amplitude) first mode, 
which justifies a monoperiodic approximation.
The amplitude of this mode is 
the largest ever observed for an SPB. The dataset used for this GEE 
application has already been shown in Figure~\ref{hd181558_spec}. 
In what follows we always assume \edit{the theoretically predicted value} 
$K = 21$.

Our first goal was to estimate $\bbeta$ for each mode $(\ell,m)$
with $\ell \le 4$, by solving the non-linear quasi-score equations. It turned 
out, however, that this was not just a technical detail of the procedure, but
was in fact a major issue. 

First, it turned out that the quasi-score function $\bU(\bbeta)$ is computationally slow to evaluate, 
with one evaluation requiring 18 time series evaluations:
6 for the moments $\mu_1$--$\mu_6$ for the working 
approximation $\bW$, and 12  for the moments $\mu_1$--$\mu_3$
for different parameters $\bbeta$ to numerically compute (with forward
differences) the derivatives in $\bD$. For this reason, prior to using (\ref{U_function}), we first determined
a good initial guess for $\hat{\bbeta}$ for the local search routine, using a rough scan of the 4D parameters space for each 
mode $(\ell,m)$ with a computationally less expensive lack-of-fit
function $g(\bbeta)$:
\begin{equation}
\label{g_function}
g(\bbeta) \equiv \sum\limits_{d=1}^3 \ \frac{1}{d}
   \ \ \sqrt[d]{\frac{1}{N} \sum\limits_{i=1}^N
   \ |y_d(t_i) - \mu_d(t_i, \bbeta)|}.
\end{equation}
The construction with the $d^{\rm th}$ root and the division by $d$ simply prevents the higher order moments from numerically dominating the lower order
moments. The sampling of the parameter space was done probabilistically
and non-uniformly. For each parameter $\beta_i$, a physical range was
determined and this range was subdivided into intervals. After each set of
10000 sampled points, each interval of each parameter $\beta_i$ was 
assigned a sampling probability according to the lowest $g(\bbeta)$ value 
recorded up to then, with the $\beta_i$ component in the corresponding
interval. The sum of probabilities over all intervals of a parameter 
$\beta_i$ was set to one. For each $(\ell,m)$ pair a total of 200,000
points was sampled. This procedure was set up to sample the more 
promising regions of the parameter space.

In Table~\ref{gmin_table} we give for the star HD181558 the lowest
$g(\bbeta)$ value recorded for each mode $(\ell,m)$.
\begin{table}
\caption{\label{gmin_table} For each $(\ell,m)$ pair, the 4D parameter space
was scanned with the lack-of-fit function $g$ defined by equation 
(\ref{g_function}) and with the dataset of the star HD181558 shown in Figure
\ref{hd181558_spec}. The minimum $g_{\rm min}$ of the $g(\bbeta)$ values for each of the $(\ell,m)$  pairs is given.}
\centering
\fbox{%
\begin{tabular}{c@{\quad}c@{\quad}c@{\quad}c@{\quad}c}
           & \multicolumn{4}{c}{$\ell$} \\
\cline{2-5}
$m$        & $1$ & $2$ & $3$ & $4$ \\
\hline
$+4\ $           &            &            &            & 11.9       \\
$+3\ $           &            &            & 7.52       & 11.0       \\
$+2\ $           &            &  6.57      & 6.71       & 11.3       \\
$+1\ $           &  4.72      &  4.74      & 5.85       & 10.8       \\
$\phantom{+}0\ $ &  6.37      &  6.37      & 6.79       & 11.6       \\
$-1\ $           &  4.79      &  6.57      & 7.05       & 10.7       \\
$-2\ $           &            &  4.68      & 6.86       & 10.5       \\
$-3\ $           &            &            & 6.92       & 11.3       \\
$-4\ $           &            &            &            & 11.3         
\end{tabular}}
\end{table}
The $(\ell,m) = (0,0)$ pair can be excluded on astrophysical grounds because 
such modes do not occur in SPBs. As can be seen, no mode $(\ell,m)$ stands out, but there are several candidate modes that 
describe the data well. Our final estimate of $\bbeta$ should take into 
account this mode uncertainty. \edit{We also note that Table~\ref{gmin_table}
is not symmetric with respect to the sign of $m$. The moments indeed behave
differently when the pulsational wave goes in the same direction as the 
rotation then when the wave goes in the opposite direction. This 
sensitivity to the
sign of $m$ was lost in the old approach (outlined in Section~\ref{currentstatus}) where one only uses the absolute value of the amplitudes.}

The 24 scans of a 4D parameter space with 200,000 points each, was a rather time
consuming but necessary task to find suitable initial guesses for 
$\hat{\bbeta}$ for the local search algorithm. We implemented  two derivative-free methods: the conjugate-direction (Powell's) method (see, e.g., Press et al., 1992, p.~420) and the Torczon (1989) simplex method. The former of which, having the best performance, was used to locate the root $\hat{\bbeta}$ of $\bU$ for all modes $(\ell,m)$. We found that these methods had much stabler performance than quasi-Newton methods, such as Newton-Raphsons, Fisher scoring, or variations to this theme. 

Even with the conjugate-direction method, the algorithm did not always converge.
The reason, as it turns out, is that the quasi-score
functions have ``false'' zeros, for example there are cases where the components 
of $\bU$ approach zero for $\sigma \rightarrow \infty$. 
Quite often, the algorithm converged to a point outside the physically relevant 
range of the parameters, even when several different initial guesses for 
$\hat{\bbeta}$ were tried. Although they did \edit{\emph{not}} occur 
for our dataset of the
star HD181558, we should mention two other possible causes of numerical
difficulties. First, it may be possible that the working approximation $\bW$ is 
not invertible, for example if $\bbeta$ approaches zero. Second, the matrix
$\bI_0$ may not be invertible, and hence no covariance matrix can be computed.
This occurs, for example, for $\alpha \rightarrow 0^{\circ}$ because
$v_e$ appears only in $v_e \sin\alpha$
in the equations, so that the third row and the third column of $\bI_0$ are zero. 
We stress, however, that the latter example is a problem of intrinsic 
non-identifiability
and is not specific for the GEE approach. One simply cannot derive the rotational
velocity if the star is looked pole-on.

\begin{figure}[h]
\centering
\begin{turn}{270}
\epsfig{file=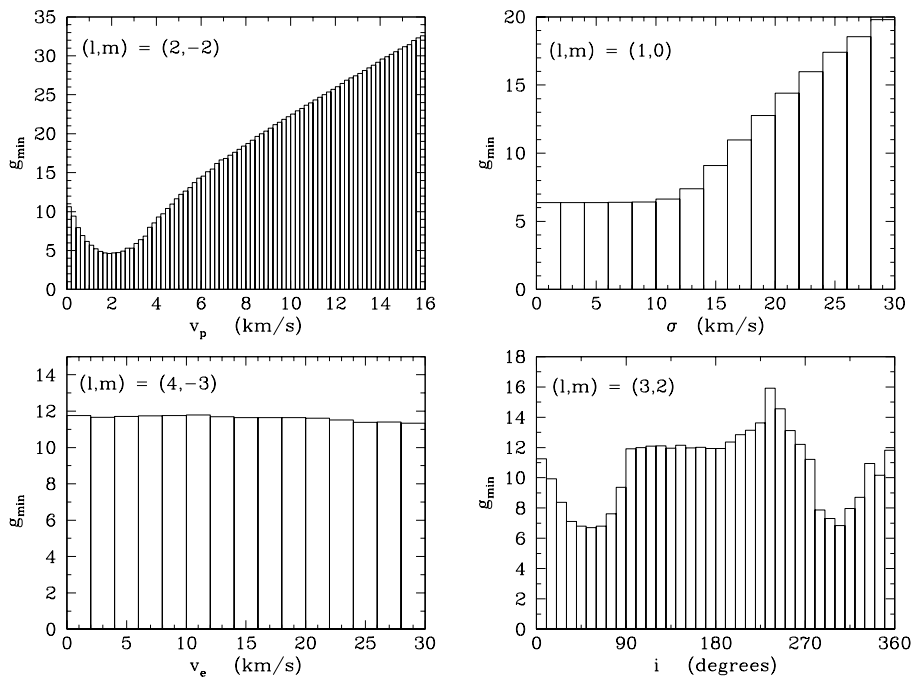,height=0.9\textwidth}
\end{turn}
\caption{Representative examples of the minimal lack-of-fit value $g_{\rm min}$ for each
sample interval of a parameter, for the star HD181558. 
We remark that although the size of the intervals for the parameter $v_p$ is fixed, the
relevant range of $v_p$ depends on the mode numbers $(\ell,m)$.
\label{mc_fig}}
\end{figure}

Making detailed 1D slices of the 4D function $\norm{\bU}$ too time
consuming, but we record the minimal lack-of-fit values 
$g_{\rm min}$ in each of the intervals of each parameter  (disregarding
the values of the other parameters).  Figure~\ref{mc_fig} shows typical examples.
Although the function $g(\bbeta)$ need not have exactly the same
behaviour as the function $\norm{\bU(\bbeta)}$ (the difference is 
similar to the well-known difference between $L_1$-norm and $L_2$-norm
minimization), we assume that the functions share many features.
We observe that the minimum in the upper left panel of Figure~\ref{mc_fig} for 
the well-fitting mode $(\ell,m) = (2,-2)$ is quite localized. This is 
much in contrast with the almost flat surface in the lower left panel 
for the badly-fitting mode $(\ell,m) = (4,-3)$. Intuitively, 
one can expect that the equivalent for the case of the $\norm{\bU}$ 
function hampers the iterations towards the minimum, and that this 
increases the chance of wandering from the physically relevant part 
of the parameter space. This is exactly what happened for this mode. 
More generally, we observe strong correlation 
between how well a mode fits the
data (with $g_{\rm min}$ as lack-of-fit value) and the chances
that the root finding algorithm does not converge. The lower
right panel shows two minima in the plot of the inclination angle $\alpha$, since we consider a full 360$^{\circ}$ range. While over such a range symmetry relations exist, these depend on $\ell$ and $m$ and taking them into account to limit the range of $\alpha$ would entail a lot of bookkeeping that can elegantly be avoided by simply considering the entire range. The feature that we quite often do not seem to find the root of $\bU$ does not necessarily contradict the theory outlined in Section \ref{GEE_theory}. 
We solve for the root of the observed $\bU$ function because we know that 
$E[\bU] = \mathbf{0}$. However, the latter is only true if the model is 
correctly specified, i.e. if $E[\bY] = \bmu(\ell,m,\bbeta)$. Therefore,
theoretically, the existence of a root in the 4D parameter space of the
continuous parameters $\bbeta$ cannot be guaranteed for a ``wrong''
$(\ell,m)$ pair, and this is exactly what we observe for badly fitting 
modes. For this reason we interpreted a non-convergence (after repeatedly
trying) as an indication that the candidate mode should be disregarded.

In Table~\ref{Powellminima}, we list the roots of the quasi-score function
for those modes for which there was convergence in the physically relevant
part of the parameter space.
\begin{table}
\caption{\label{Powellminima}
Roots of the quasi-score functions for those modes where there was
convergence in the physically relevant part of the parameter space, for the
star HD181558. The values between brackets are the standard errors obtained
with the sandwich estimator (\ref{cov_beta}). 
$G^2$ is the {lack-of-fit value} of the mode as defined by Eq.~(\ref{G_weight}). 
$v_p$, $\sigma$ and $v_e$ are expressed in km/s, and the inclination angle $\alpha$ in 
degrees.}
\centering
\fbox{%
\begin{tabular}{c@{\qquad}c@{\qquad}c@{\qquad}c@{\qquad}c@{\qquad}c@{\qquad}c}
$(\ell,m)$ & $\norm{\bU}_{\rm min}$ & $G^2$ & $\widehat{v}_p$ & $\widehat{\sigma}$ & $\widehat{v}_e$ & $\widehat{\alpha}$ \\
\hline
(1,0)  & 0.15            & 2.7  & 2 (1)      & 9.0 (0.9) & 13 (19)    & 320 (48)   \\
(1,1)  & 5.1\ 10$^{-24}$ & 0.63 & 2.01 (0.08) & 6.3 (0.2) & 15.2 (0.7) & 117 (2)   \\
(1,-1) & 1.1\ 10$^{-5}$  & 1.1  & 4.0 (0.2)   & 4.2 (0.6) & 25 (2)     & 336 (1)   \\
(2,0)  & 0.10            & 3.1  & 0.9 (0.4)   & 7 (2)     & 30 (24)    & 331 (13)  \\
(2,1)  & 1.7             & 0.61 & 1.7 (0.1)   & 3.8 (0.8) & 16 (1)     & 71 (1)    \\
(2,2)  & 0.014           & 2.4  & 1 (1)       & 10 (2)    & 0.4 (29)   & 270 (360) \\
(2,-2) & 0.0032          & 0.72 & 1.62 (0.06) & 4.3 (0.4) & 17.6 (0.7) & 129 (1)   \\
(3,1)  & 1.1\ 10$^{-25}$ & 2.3  & 1.00 (0.04) & 6.8 (0.7) & 18 (2)     & 145 (7)   \\
(3,2)  & 0.40            & 3.1  & 1.1 (0.2)   & 6 (3)     & 17 (10)    & 49 (23)   \\
(3,-1) & 3.5             & 7.0  & 1.3 (0.3)   & 3 (12)    & 7 (12)     & 189 (5)   \\
(4,0)  & 2.7             & 11   & 0.4 (0.3)   & 8 (12)    & 49 (188)   & 19 (96)   \\
(4,-4) & 0.047           & 8.7  & 0.6 (3)     & 7 (24)    & 20 (87)    & 295 (30)
\end{tabular}}
\end{table}
The closeness of  $\norm{\bU}$ to zero varies from mode to mode. In cases where this value fails to be small, we checked this is not due to premature convergence, since 
restarting the algorithm at the point where it stopped, did not further decrease 
$\norm{\bU}$. One possible explanation might be that, for some of the
solutions, the algorithm has converged to a local minimum, such as for
the modes $(\ell,m) = (3,-1)$ and $(4,0)$.

\begin{figure}[h]
\centering
\begin{turn}{270}
\epsfig{file=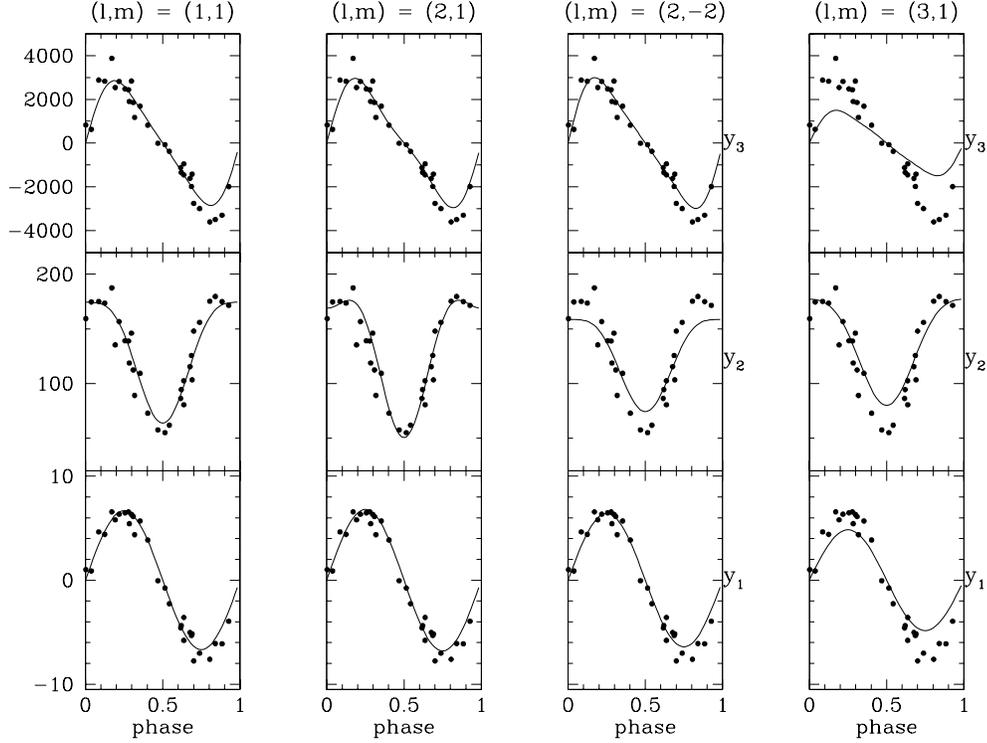,height=\textwidth}
\end{turn}
\caption{Theoretical models (solid lines) of the observed moments (bullets) for the
three best fitting modes $(\ell,m) = (1,1)$, $(\ell,m) = (2,1)$, and 
$(\ell,m) = (2,-2)$, plus the poorer fitting mode $(\ell,m) = (3,1)$, 
with $G^2$ as a lack-of-fit function. The theoretical models were obtained 
with the model parameters obtained with the GEE method.
The first, the second and the third row are for the first moment $y_1$ (km/s), 
the second moment $y_2$ (km$^2$/s$^2$) and the third moment $y_3$ (km$^3$/s$^3$) 
respectively.
The moments are shown as a function of the phase. Note that the models of the different promising modes differ 
mainly for the second moment.\label{fits}}
\end{figure}

In Figure~\ref{fits} we show fits for the three best fitting modes, with 
the function $G^2$ (see Eq.~\ref{G_weight}) as a lack-of-fit.
As mentioned before, there is not just one, but several modes that can fit 
the observed data quite well. 
The lack-of-fit values in Table~\ref{gmin_table} provide us with 
an indication of the relative merits of wave number choice $(\ell,m)$. Of course, at this 
point we lack knowledge about the reference distribution of these values, 
unlike in classical fit statistics (e.g.,  likelihood-ratio based). However, 
similar instances exist in both a frequentist (e.g., Akaike Information Criterion) 
and a Bayesian context (e.g., Bayes factors). Nevertheless, we assert that these 
numbers, especially when supported by careful graphical inspection, are useful 
to narrow down substantially our uncertainty about the wavenumbers, 
in spite of an intrinsically complicated modelling endeavor. To this end, 
the last column in Figure~\ref{fits} displays mode $(\ell,m)=(3,1)$ with a 
substantially worse fit than the one in the first three columns of the same figure. 

We used the modes in Table \ref{Powellminima}, 
to compute the \emph{weighted} mean $\tilde{\bbeta}$ and its standard error, with 
Eqs.~(\ref{tildebeta}) and (\ref{cov_tildebeta}). The results, including 
the intra-mode and inter-mode variance, are given in Table 
\ref{weightedmean_12modes}.
\begin{table}
\caption{The weighted mean over all 12 modes in Table~\ref{Powellminima}, computed
with Eqs.~(\ref{tildebeta}) and (\ref{cov_tildebeta}). The values mentioned 
between brackets are standard errors. The intra-mode variance and inter-mode
variance are computed with respectively the first and the second term of
(\ref{cov_tildebeta}).  $\tilde{v}_p$, $\tilde{\sigma}$, $\tilde{v}_e$ are
expressed in km/s, and the inclination angle $\tilde{\alpha}$ is expressed in 
degrees. \label{weightedmean_12modes}}
\centering
\fbox{%
\begin{tabular}{cccc}
$\tilde{\beta}_i$ & Weighted Mean & Intra-mode Variance & Inter-mode Variance \\
\hline
$\tilde{v}_p$    & 1.8 (1.0) & 0.33 & 0.74 \\
$\tilde{\sigma}$ & 5.5 (4.1) & 14   & 3.1  \\
$\tilde{v}_e$    & 17 (26)   & 612  & 46   \\
$\tilde{\alpha}$     & 164 (132) & 7300 & 10079
\end{tabular}}
\end{table}
In the specific case of HD181558, one could argue that the modes with $\ell = 3$ 
and $\ell = 4$ can be disregarded on astrophysical grounds. The reason is that
these modes would require a very large oscillation amplitude at the surface 
of the star to cause the large observed amplitude of the first moment $y_1$. For
this reason, we also computed $\bbeta$ with the $\ell =1$ and $\ell = 2$ modes
of Table \ref{Powellminima} only. The results are listed in Table~\ref{weightedmean_7modes}.
\begin{table}
\caption{The same information as in Table \ref{weightedmean_12modes} is shown,
except that the mean is computed over those 7 modes in Table 
\ref{Powellminima} with $\ell = 1$ and $\ell = 2$.
\label{weightedmean_7modes}}
\centering
\fbox{%
\begin{tabular}{cccc}
$\tilde{\beta}_i$ & Weighted Mean & Intra-mode Variance & Inter-mode Variance \\
\hline
$\tilde{v}_p$    & 2.0 (1.0) & 0.19 & 0.71 \\
$\tilde{\sigma}$ & 5.3 (2.0) & 0.82 & 3.1 \\
$\tilde{v}_e$    & 16 (12)   & 100  & 37   \\
$\tilde{\alpha}$     & 170 (137) & 8342 & 10450 
\end{tabular}}
\end{table}
With Tables \ref{weightedmean_12modes} and \ref{weightedmean_7modes} we 
achieve the goal of this application of the revised version of the moment 
method: we have obtained a best guess 
for the continuous parameters and their standard errors, where we took into 
account the mode uncertainty. An important result is that the uncertainties 
of the parameters can be large, in fact larger than we expected. Especially the
rotational velocity $v_e$ cannot be estimated precisely. The large inter-mode
uncertainty of the inclination angle $\alpha$ is not surprising since the inclination
angle is known to be largely dependent on the mode numbers $(\ell,m)$. 
We note that the values for the weighted means do not change 
much by excluding the $\ell = 3$ and $\ell = 4$ modes. The reason is that 
the latter modes have a lower weight anyway, as can be seen from the $G^2$ 
values in Table \ref{Powellminima}. 

The fact that the standard errors for the continuous parameters are large turns out not to be specific for the star HD181558. We applied our method to several artificial datasets, and we obtained similar results. Hence, we conclude that estimates of the continuous parameters generally can be quite uncertain.


\section{Summary and Conclusions}

We made a first but arguably important step to develop a statistical 
formalism for
the moment method. In this first stage we aimed to incorporate estimates 
of the uncertainties of the continuous parameters. Because of the 
many difficulties to overcome, this was never done for the 
moment method, nor for any other mode identification technique.

We found that, in the specific case of the moment method, the method of 
least-squares does not give consistent estimates of the continuous
parameters and we resort to the GEE method (Liang and Zeger 1986). This method requires a working approximation of the covariance matrix of the 3 responses, based on the higher
theoretical moments. Note that the higher moments, known to be imprecise, are {\em not\/} used in the actual model. An important source of uncertainty is the fact that often not just one but several candidate modes can describe the data. We set up a separate procedure to weight each mode and to compute a weighted mean over all modes of the parameter vector and its uncertainty. To compute the latter we introduced the intra-mode and the inter-mode
uncertainty. 

Subsequently, we applied our procedure to the SPB star HD181558,
from which we learned the strong and the weak points of our method.
We found out that solving the estimating equations is a 
computationally demanding and tedious task: convergence of the algorithm 
was not evident, despite the fact that we experimented with several robust 
local root finding methods, of which we selected the method of conjugate 
directions as the most efficient one. On the other hand, we also proposed
a new lack-of-fit function to scan the parameter space to obtain good
initial guesses for the local search method. This lack-of-fit function
proved to be very useful on its own, as it works at least as well as the 
old discriminant (\ref{discriminant}) and allows in addition to discriminate
between positive and negative azimuthal numbers which was one of the
shortcomings of the previous discriminant. Our strategy
to scan the parameter space also proved that there are several modes 
that can explain the dataset of HD181558. Only taking into account
the very best fitting one, would therefore not be useful.

This is why we retained 12 modes as candidate modes for which an estimate of the continuous
parameters $\hat{\bbeta}$ can be computed, and used these estimates
to obtain a best guess $\tilde{\bbeta}$ for the continuous parameters
plus their uncertainties, taking into account the mode uncertainty.
Doing so, we discovered that the parameter uncertainties can be large,
a result which was moreover confirmed in the case of artificial 
datasets.

{Prior to our study, such large uncertainties were not anticipated.
On the contrary, one sometimes assumed
the uncertainties to be quite small in order to be able to apply a two-stage approach in the multiperiodic case. In such an approach} the inclination angle 
$\alpha$ and the rotational velocity $v_e$ {are} determined with the 
dominant mode, and subsequently fixed while determining the mode parameters 
of the other modes, to have the dimension of the parameter space reduced. 
Our {new} results show that such an approach can be very dangerous: 
in the case of HD181558 it can hardly be justified because of the large 
uncertainty on $\alpha$.

\edit{The method we outlined in this paper is the very first attempt
to develop a statistical formalism for the moment method.}
Even though there is undoubtedly room for additional work before our proposed method can be deemed widely applicable, we conclude it makes an important first step in our understanding of the uncertainties and usefulness of the continuous parameters, estimated with the moment method. It furthermore underscores that some conclusions reached in the past need to be revisited.
 
We conclude with several possible future improvements.
First, it may be worth to investigate alternative parameterizations
$\bbeta'$ which, while mathematically equivalent to the original one, may improve upon the convergence and robustness properties
of the algorithm. We already experimented, for example, with using 
$v_e \sin\alpha$ instead 
of $v_e$, and with using $v_p \sin\alpha$ and $v_p \cos\alpha$ instead of 
$v_p$ and $\alpha$. Second, it would be useful to extend the formalism 
to include the uncertainty on $K$. Third, it might also be interesting
to use several spectral lines at the same time to improve the statistics.
Including multiple modes might also improve the convergence 
properties. Although this would imply 3 more parameters 
$(l_2, m_2, v_{p,2})$ per mode, multiple modes would set more 
stringent restrictions on the inclination angle $\alpha$ which has a 
significant impact on the estimation of the other parameters.
To further validate all of the above, more simulations are necessary. In addition, such simulations can also clarify what impact the number and the signal 
to noise ratio of the observational spectral lines has on the performance
of the algorithm.


\section*{References}

Aerts, C., De Pauw, M. and Waelkens, C. (1992) Mode identification of pulsating 
stars from line profile variations with the moment method. An example - The 
Beta Cephei star Delta Ceti, \emph{Astronomy and Astrophysics}, \textbf{266}, 294-306. \\

Aerts, C. (1996) Mode identification of pulsating stars from line-profile 
variations with the moment method: a more accurate discriminant, 
\emph{Astronomy and Astrophysics}, \textbf{314}, 115-122 \\

Aerts, C., De Cat, P., Cuypers, J., Becker, S.R., Mathias, P., De Mey, K., 
Gillet, D. and Waelkens, C. (1998) Evidence for binarity and multiperiodicity 
in the beta Cephei star beta Crucis, \emph{Astronomy and Astrophysics}, \textbf{329},
137-146 \\

Aerts, C. and Kaye, A.B. (2001) A Spectroscopic Analysis of the gamma Doradus Star 
HD 207223 = HR 8330, \emph{The Astrophysical Journal}, \textbf{553}, Issue 2, 814-822 \\

Balona, L.A. (1986) Mode identification from line profile variations, 
\emph{Monthly Notices of the Royal Astronomical Society}, \textbf{220}, 647-656 \\

Chadid, M., De Ridder, J., Aerts, C. and Mathias, P. (2001) 20 CVn: A monoperiodic 
radially pulsating delta Scuti star, \emph{Astronomy and Astrophysics}, \textbf{375}, 
113-121 \\

De Cat, P. and Aerts, C. (2002) A study of bright southern slowly pulsating B
stars. II. The intrinsic frequencies, \emph{Astronomy and Astrophysics}, 
\textbf{393}, 965-981 \\

Diggle, P.J., Heagerty, P., Liang, K.Y. and Zeger, S.L. (2002)
\emph{Analysis of Longitudinal Data}, Oxford University Press, Oxford \\

Liang, K.Y. and Zeger, S. (1986) Longitudinal data analysis using generalized
linear models, \emph{Biometrika}, \textbf{73}, 13 \\

Press, W.H., Teukolsky, S.A., Vetterling, W.T. and Flannery, B.P. (1992) 
\emph{Numerical Recipes in C. The art of scientific computing.} (2nd ed.), 
Cambridge University Press \\

Seber, G.A.F. and Wild, C.J. (1989) \emph{Nonlinear Regression}, Wiley series in
probability and mathematical statistics, John Wiley \& Sons, Inc. \\

Torczon, V.J. (1989) Multi-directional search: a direct search algorithm
for parallel machines, PhD Thesis, Rice University, Houston (Texas, U.S.A.) \\

Uytterhoeven, K., Aerts, C., De Cat, P., De Mey, K., Telting, J.H., Schrijvers, C.,
De Ridder, J., Daems, K., Meeus, G. and Waelkens, C. (2001) Line-profile variations 
of the double-lined spectroscopic binary kappa Scorpii, \emph{Astronomy and 
Astrophysics}, \textbf{371}, 1035-1047 \\

Zeger, S.L., Liang, K.Y. (1986) Longitudinal data anlaysis for discrete and
continuous outcomes, \emph{Biometrics}, \textbf{42}, 121-130 \\

\end{document}